\documentclass[12pt]{article}
\usepackage{amsmath}
\usepackage{graphicx}
\usepackage{enumerate}
\usepackage{natbib}
\usepackage{array}
\usepackage{url} 

\usepackage{blindtext}
\usepackage{floatrow}
\usepackage[toc,page]{appendix}
\newfloatcommand{capbtabbox}{table}[][\FBwidth]

\usepackage{tabularx}
\usepackage{booktabs}
\usepackage{xcolor}
\usepackage{algorithm}
\usepackage[noend]{algpseudocode}

\usepackage{multirow}
\usepackage{amssymb}
\usepackage{pifont}

\usepackage[T1]{fontenc}
\usepackage[utf8]{inputenc}
\usepackage[font=small,labelfont=bf]{caption}
\usepackage{lscape} 
\usepackage{changepage}
\usepackage{amsthm}

\usepackage[utf8]{inputenc}
\usepackage{natbib}
\setcitestyle{round, numbers, comma, sort&compress}
\usepackage{hyperref}
\usepackage{subcaption}
\usepackage{graphicx}
\newtheorem{theorem}{Theorem}[section]
\newtheorem{definition}{Definition}[section]

\newcommand{\blind}{1}
\makeatletter
\newcommand{\vast}{\bBigg@{3.5}}
\newcommand{\bm}[1]{\mbox{\boldmath$ #1 $\unboldmath}}
\newcolumntype{C}[1]{>{\centering\let\newline\\\arraybackslash\hspace{0pt}}m{#1}}

\makeatother

\newtheorem{innercustomgeneric}{\customgenericname}
\providecommand{\customgenericname}{}
\newcommand{\newcustomtheorem}[2]{%
  \newenvironment{#1}[1]
  {%
   \renewcommand\customgenericname{#2}%
   \renewcommand\theinnercustomgeneric{##1}%
   \innercustomgeneric
  }
  {\endinnercustomgeneric}
}

\newtheorem*{assumption*}{\assumptionnumber}
\providecommand{\assumptionnumber}{}


\newcustomtheorem{customthm}{Theorem}

\begin{document}

\def\spacingset#1{\renewcommand{\baselinestretch}%
{#1}\small\normalsize} \spacingset{1}


\if1\blind
{
  \title{\bf Deep P-Spline: Theory, Fast Tuning, and Application}
  \author{Noah Yi-Ting Hung$^{1}$, Li-Hsiang Lin$^{1}$, and Vince D. Calhoun$^{2}$ \\
  $^{1}$Department of Mathematics and Statistics,\\
  Georgia State University, Atlanta, GA, 30303\\
  $^{2}$Tri-institutional Center for Translational Research in Neuroimaging and\\
  Data Science, Georgia State University, Georgia Institute of Technology,\\
  Emory University, Atlanta, GA, 30303}
  
  \maketitle
} \fi

\if0\blind
{
  \bigskip
  \bigskip
  \bigskip
  \begin{center}
    {\LARGE\bf Deep P-Spline: Theory, Fast Tuning, and Application}
\end{center}
  \medskip
} \fi

\bigskip
\begin{abstract}
Deep neural networks (DNNs) have been widely applied to solve real-world regression problems. However, selecting optimal network structures remains a significant challenge. This study addresses this issue by linking neuron selection in DNNs to knot placement in basis expansion techniques. We introduce a difference penalty that automates knot selection, thereby simplifying the complexities of neuron selection. We name this method Deep P-Spline (DPS). This approach extends the class of models considered in conventional DNN modeling and forms the basis for a latent variable modeling framework using the Expectation-Conditional Maximization (ECM) algorithm for efficient network structure tuning with theoretical guarantees. From a nonparametric regression perspective, DPS is proven to overcome the curse of dimensionality, enabling the effective handling of datasets with a large number of input variables—a scenario where conventional nonparametric regression methods typically underperform. This capability motivates the application of the proposed methodology to computer experiments and image data analyses, where the associated regression problems involving numerous inputs are common. Numerical results validate the effectiveness of the model, underscoring its potential for advanced nonlinear regression tasks.
\end{abstract}

\noindent%
{\it Keywords:} Smoothing and nonparametric regression; Surrogate modeling; Difference penalty; Feature selection;  Regularization
\vfill

\newpage
\spacingset{1.9} 
\section{Introduction}


Deep neural networks (DNNs), have gained immense popularity across various scientific fields due to their flexibility and ability to model complex  regression relationships \citep{lecun2015deep, krizhevsky2012imagenet,ioffe2015batch}. However, a notable challenge in DNNs is the selection of an appropriate network structure, such as the number of neurons and layers, which is crucial for balancing model complexity and performance \citep{huang2018data,murdock2016blockout,diaz2017effective,anders1999model}. In this study, basis expansion techniques \citep{hastie2009elements, fan2020statistical} with a novel penalty for DNNs modeling, based on difference penalty, is introduced to address the challenge. While difference penalties have been extensively studied in conventional nonparametric smoothing models \citep{eilers1996flexible, li2008asymptotics}, their extension in the deep learning realm remains under explored.


Efficiently selecting a neural network structure, including the optimal number of neurons and layers, is a complex and computationally demanding problem. An inappropriate network structure significantly influences both the model performance and computational efficiency. Traditional selection methods are cross-validation- or grid-search- based method \citep{feurer2019hyperparameter, browne2000cross}, often computationally prohibitive with limited insights into the underlying selection mechanisms. We present a new perspective for the network selection problem by establishing a connection between the selection of neuron numbers and the selection of knot locations, the positions for placing a spline expansion as a neuron function. This perspective extends previous statistical frameworks to DNN structure, allowing for improved computational efficiency in network structure selection. Specifically, we leverage the difference penalty to create an automated knot selection procedure, effectively circumventing the difficulties associated with neuron selection. This approach also lays the foundation for a novel latent variable modeling framework, which incorporates an Expectation Conditional Maximization (ECM)-based algorithm \citep{meng1993maximum} for efficient network structure tuning. By iteratively optimizing a latent variable model that integrates the difference penalty within the DNN architecture, the ECM algorithm enables precise and efficient network structure selection. Advantages of the network structure selection method will be demonstrated will explored, including  its efficiency and ability to overcome model selection problem observed from Double Decent phenomenon.

The purpose of introducing a penalty in our deep neural network (DNN) framework differs significantly from other penalty-based DNN methods. For instance, a Lasso penalty has been used in DNNs for variable selection to identify important input features \citep{lemhadri2021lassonet}. These penalized DNN method considers constrains placed on the DNN coefficients. A summary of previous penalized or coefficient constraints methods are given in Table \ref{tab:PenalizedDNN}. Among these methods, the total variation (TV) penalty, which is also related to the differentials properties of the DNN models, has been applied to DNNs modeling to encourage sparsity in the second derivative of the model, drawing a connection to the spline representation theorem \citep{wahba1990spline}. This TV penalty closely resembles the roughness penalty, which involves integrating the square of the second derivative—a technique commonly used in spline modeling history. Both TV and roughness penalties are continuous, leading to a strong coupling between the penalty order and the degree of the spline basis functions \citep{eilers2021practical}. In contrast, our proposed difference penalty does not impose this coupling, providing greater flexibility in model design.

\begin{table}[h!] 
\resizebox{16cm}{!}{
\begin{tabular}{|c|c|c|c|}
\hline
Literature & {Neuron Function} & Convergence Rate & Regularization Type \\ \hline
\citep{lemhadri2021lassonet}   & Linear & X &    $L_{1}$ norm    \\ \hline
\citep{bohra2020learning}   & Spline & X & Total Variation Norm\\ \hline
\citep{parhi2021banach} & Spline & X & Total Variation Norm\\ \hline
\citep{sun2022consistent} & Linear & O & Parameter constraints\\ \hline
\end{tabular}}
\caption{A summary of recent literature on DNNs with Penalty or network coefficients constraint.}
\label{tab:PenalizedDNN}
\end{table}

This paper also addresses the theoretical guarantees of the proposed framework. While deep neural networks (DNNs) have achieved substantial practical success, their theoretical underpinnings—particularly regarding convergence rates—remain an active area of research. Recent studies suggest that convergence rates for DNNs are highly dependent on the number of neurons and layers in the network \citep{kohler2021rate}, highlighting the critical role of structural design. We extend these insights by demonstrating that, under the penalized framework, the primary factor affecting convergence rates is the effective dimension of each layer. This distinction provides a more nuanced understanding of the impact of the penalized network structure on model performance and offers a theoretical basis for our approach.

Moreover, our convergence rate analysis reveals an intriguing result: when the proposed method is treated as a nonparametric regression model, it effectively circumvents the curse of dimensionality—a common limitation in conventional nonparametric regression \citep{donoho2000high}. This finding significantly broadens the potential applications of the proposed method, making it highly effective for nonlinear regression problems with numerous input variables. By overcoming the curse of dimensionality, this penalized DNN framework efficiently detect complex nonlinear regression associations with many inputs. Along this direction, we apply the proposed method as a surrogate modeling technique in computer experiments and as a classifier for image data, showcasing its utility in diverse applications.

The remainder of the paper is organized as follows. In Section 2, we introduce the new framework of using basis expansion with difference penalty on DNN. Section 3 presents a fast notwork structure selection method for use with theoretical guarantees. Intensive simulation studies are demonstrated in Section 4. Section 5  explores the theoretical properties of the proposed framework. In Section 6, we demonstrate the effectiveness of our approach through two major applications in surrogate modeling and image data analysis. Finally, concludes  and potential future research directions are discussed in Section 7.

\section{Proposed Method}\label{sec:def}

A Deep Neural Network consists of multiple hidden layers of neurons between the input variables $\mathbf{x} \in \mathbb{R}^d$ and the output variables $y \in \mathbb{R}$. Let the network structure be denoted by $\mathcal{N} = (L, \mathbf{p})$, where $\mathbf{p} = \{p_{(1)}, \ldots, p_{(L)}\} \in \mathbb{N}^{L}$ specifies the number of neurons in each layer, and $L$ represents the number of layers. The first layer is called {\it input layer} because its connection to the input variables, and the neuron values of the $(L-1)$-th hidden layer are connected to the output variable $y$ through another function, such as a linear combination of all the neuron values in this layer, so the ${p}_{(L)} = 1$ and $L$-th layer is called the {\it output layer}. For layers before the output layer are called {\it hidden layers}, their neuron values are usually further applied by an activation function $\phi_{(\ell)}: \mathbb{R}^{p_{(\ell)}} \rightarrow \mathbb{R}^{p_{(\ell)}}$ for $\ell = 1, 2, \cdots, L-1$ to introduce extra function for better training via nodes permutation, regression sign changes, weight rescaling, and nonlinearity consideration \citep{sun2022consistent}. The common DNN models consider the value of a neuron in a hidden layer is a weighted linear combination of outputs from the previous layer, followed by an activation function such as the Rectified Linear Unit (ReLU) \citep{nair2010rectified}. Specially, if the observed dataset is expressed as $\{(\mathbf{x}_{i}, y_{i}): \mathbf{x}_{i} \in \mathbb{R}^{d} \mbox{ and } y_{i} \in \mathbb{R}\}$, then the DNN models can be expressed as
\begin{eqnarray}\label{eq:Ori-DNN}
{\mathbf{h}_{(1)}\mathbf{x}_{i}} & = & \phi_{(1)}\left(\mathbf{W}_{(1)}\left(\mathbf{x}_i\right)\right) \nonumber \\
& \vdots & \nonumber \\
\mathbf{h}_{(L-1)}\left(\mathbf{h}_{(L-2)}\right)  & = & \phi_{(L-1)}(\mathbf{W}_{(L-1)} \mathbf{h}_{(L-2)}), \label{eq: DNN}\\
 h_{(L)}(\mathbf{h}_{(L-1)}) & = & \mathbf{w}_{(L)} \mathbf{h}_{(L-1)} \nonumber\\
{y_{i}} & = & g\left(h_{(L)}(\mathbf{h}_{(L-1)})\right) + \epsilon_{i} \nonumber 
\end{eqnarray}
\noindent  where $\epsilon_{i}$ is the observed noise, $\mathbf{W}_\ell \in \mathbb{R}^{p_{(\ell)} \times p_{(\ell-1)}}$ for $\ell = 1, \cdots, L-1$ is the weight matrix connecting neurons in $(\ell-1)$-th layer to those in $\ell$-th layer for $\ell = 1, \cdots, L-1$ and $\mathbf{w}_{(L)}$ is the linear combination coefficients of the output layer. In addition, $g(\cdot)$ is usually determined with respect to different tasks as
\begin{eqnarray}
    g(\cdot) = \left\{ \begin{array}{rcl}
    &I(\cdot) & \mbox{for regression problem} \\ 
    &\text{softmax}(\cdot) & \mbox{for classification problem.} 
\end{array}\right. \nonumber
\end{eqnarray}

\subsection{Equivalence between Neuron Count and Knot Selection in Spline Expansion}
The model class of each neuron function in model (\ref{eq: DNN}) is piecewise linear and globally continuous \citep{elbrachter2019deep, lu2020universal}. According to spline theory \citep{balestriero2018spline}, any continuous function can be closely approximated by a linear combination of higher-order spline functions. This motivate us to consider the model class of the neuron functions as the combinations of spline bases, which extends the model class to higher-order continuous function instead of linear class.

The approximation of functions by a linear combination of higher-order splines also addresses the neuron-selection problem in each layer of a neural network. An intuition is because the approximation properties of splines suggest a relationship between the number of neurons in hidden layer of a linear structure DNN and the placement of knots in a spline. We summarize this connection in the following theorem.

\begin{figure}[h!]
    \centering
    \includegraphics[width=1.0\linewidth]{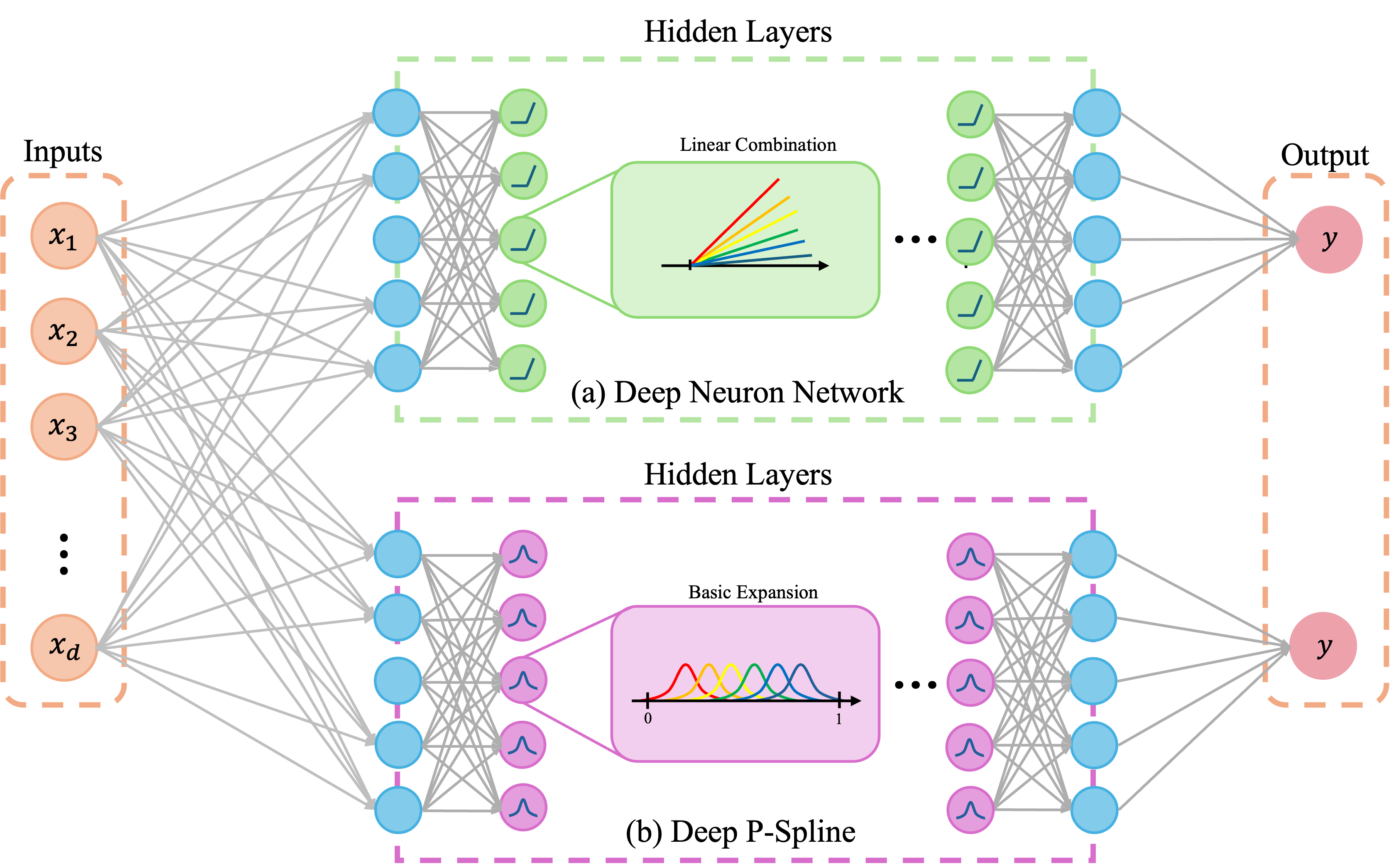}
    \caption{(a) DNN and (b) the proposed DPS structure}
    \label{fig:DPS-Structure}
\end{figure}

\begin{theorem}\label{thm: knotselection}
A weighted sum of the outputs of each hidden layer in a DNN can be well-approximated by a B-spline basis function, provided the number of knots is sufficient.
\end{theorem}
\noindent The proof is provided in Appendix A of Supplemental Material. This result indicates that a single spline neuron function with adequate knots is a linear combination of several neurons. As a result, the selection of neuron numbers becomes a research problem about selecting knots locations and numbers, which can be solved efficiently by considering a difference penalty as illustrated in next subsection. Traditionally, determining the optimal number of neurons is challenging and computationally expensive, often involving cross-validation or exhaustive grid search. In contrast, knot selection in spline functions has been extensively studied, with efficient and systematic methods available to simplify this process. Although this result stems from standard spline approximation theory, its use in designing efficient DNNs is novel and serves as a key motivation for our work.  These promising motivation encourages us to consider and study the basis expansion on $\ell$-th layer instead of a linear mapping for $\ell = 2,\cdots, L-1$ in (\ref{eq: DNN}):
\begin{eqnarray}\label{eq: nonlin_DL}
        \mathbf{h}_{(1)}(\mathbf{x}_{i}) & = & \phi_{(1)}(\mathbf{W}_{1}\mathbf{x}_{i}) = (h_{1, (1)}, h_{2, (1)}, \cdots, h_{p_{(1)}, (1)})  \nonumber \\
         \mathbf{h}_{(\ell)}(\mathbf{h}_{(\ell-1)}) & = & \phi_{(\ell)}\left(\sum^{N_{(\ell)}}_{k=1}w_{1k, (\ell)} B_{k, (\ell)}( \mathbf{h}_{(\ell-1)}), \cdots, \sum^{N_{(\ell)}}_{k=1}w_{p_{(\ell)}k, (\ell)} B_{k,(\ell)}( \mathbf{h}_{(\ell-1)})\right) \nonumber\\
        & \equiv & \left(h_{1, (\ell)}( \mathbf{h}_{(\ell-1)}), \cdots,  h_{p_{(\ell)},(\ell)}( \mathbf{h}_{(\ell-1)})\right) \mbox{ for } \ell = 2,\cdots, L-1 \\
        h_{(L)}  & = & \mathbf{w}_{(L)}\mathbf{h}_{(L-1)} \mbox{ and }  y_i=  g(h_{(L)}) \nonumber
\end{eqnarray}
where $B_{j,(\ell)}$ is a $D$-degree B-Spline basis with chosen knots locations for $\ell = 2, \cdots, L-1$, $N_{(\ell)}$ is the knot numbers for a spline expansion in a neuron function, and $w_{jk, (\ell)}$ is the spline combination coefficient for $j$-th neuron in $\ell$-th layer for $j = 1, \cdots, p_{(\ell)}$ and $\ell = 2, ..., L-1$, and $g(\cdot)$ is the output layer function depending on the purpose of using the DNN, as discussed in (\ref{eq: DNN}). A demonstration to compare Model (\ref{eq: DNN}) and (\ref{eq: nonlin_DL}) is provided in Figure \ref{fig:DPS-Structure}.
Extending from conventional DNN model (\ref{eq: DNN}),  model (\ref{eq: nonlin_DL}) further utilizes the order of values from the neurons in previous layers to have flexibility in introducing nonlinearity. This, the selection of neuron numbers becomes a problem of the selection of knot, which can be solved efficiently as discussed in the following subsection. 


\subsection{Difference Penalty for Automated Knot Selection}

To automate the knot selection, we leverage the order of the knots in Figure \ref{fig:DPS-Structure} and apply a difference penalty \citep{eilers2010splines}. The difference penalty encourages similar effects among adjacent knots, reducing the need for manual placement. This approach simplifies the training process and ensures smoothness in the fitted function. Following the notation in model (\ref{eq: nonlin_DL}), we propose the following objective function with layer-wise difference penalty for training the proposed model:


\begin{equation}\label{eq: obj}
\sum^{n}_{i=1} \left(y_{i} - f_{\mathbf{W}_{(1)}, \cdots, \mathbf{W}_{(L-1)},{\mathbf{w}_{(L)}}}(\mathbf{x}_{i}) \right)^2 + \sum_{\ell=2}^{L-1} \lambda_{(\ell)} \left( \sum^{p_{(\ell)}}_{j=1}\left| \left| \mathbf{D}_{(\ell), r} \boldsymbol{\omega}_{j,(\ell)} \right| \right|^2\right),
\end{equation}
where $f_{\mathbf{W}_{(1)}, \cdots, \mathbf{W}_{(L-1)},\mathbf{w}_{(L)}}(\mathbf{x}_{i}) = h_{(L)}\left(\mathbf{h}_{(L-1)}\left(\cdots \mathbf{h}_{(1)}\left(\mathbf{x}_{i}\right)\right)\right)$ from model \ref{eq: nonlin_DL}, the $(j, k)$-th element $\omega_{jk, (\ell)}$ of $\ell$-th layer combination coefficient matrix $\mathbf{W}_{(\ell)}$ for $j = 1, \cdots, p_{(\ell)}$, $k = 1, \cdots, N_{(\ell)}$, and $\ell = 1, \cdots, L-1$ are defined in (\ref{eq: nonlin_DL}) as well as functions $\{\mathbf{h}_{(\ell)}\}^{L-1}_{\ell = 1}$, and $ \boldsymbol{\omega}_{j,(\ell)}$ is the $j$-th row of $\mathbf{W}_{(\ell)}$. In (\ref{eq: obj}), $\{\lambda_{(\ell)}\}^{L-1}_{\ell = 2}$ are penalty parameters controlling regularization in each hidden layer $\ell = 2, \cdots, L-1$, and $\mathbf{D}_{(\ell), r} \in \mathbb{R}^{(N_{(\ell)}-r) \times N_{(\ell)}}$ is the $r$-th order difference matrix; for example, the first two orders are the most commonly used and defined as

\[
\mathbf{D}_{(\ell), 1}  =
\begin{bmatrix}
-1 & 1 & 0 & \cdots & 0 \\
0 & -1 & 1 & \cdots & 0 \\
& \ddots & & \\
0 & \cdots & 0 & -1 & 1
\end{bmatrix};
\quad
\mathbf{D}_{(\ell), 2}  =
\begin{bmatrix}
-1 & 2 & -1 & 0 & \cdots & 0 \\
0 & -1 & 2 & -1 & \cdots & 0 \\
& \ddots & & \\
0 & \cdots & 0 & -1 & 2 & -1
\end{bmatrix}.
\]
The penalty term $\sum_{\ell=2}^{L-1} \lambda_{(\ell)} \left( \sum^{p_{(\ell)}}_{j=1}\left| \left| \mathbf{D}_{(\ell), r} \boldsymbol{\omega}_{j,(\ell)} \right| \right|^2\right)$ enforces smoothness across the layers, reducing overfitting and promoting a smoother approximation of the underlying function.  We name the proposed model (\ref{eq: nonlin_DL}) with objective function (\ref{eq: obj}) by \textbf{Deep Penalty Splines (DPS)} method.\par

To optimize objective function (\ref{eq: obj}), we compute the gradient with respect to the weights $\mathbf{W}_{(\ell)}$ in each layer. The gradient consists of two components: (1) the gradient of the loss term $\sum^{n}_{i=1} \left(y_{i} - f_{\mathbf{W}_{(1)}, \cdots, \mathbf{W}_{(L-1)},\mathbf{w}_{(L)}}(\mathbf{x}_{i}) \right)^2$, and (2) the gradient of the penalty term. The gradient can be calculated conveniently in modern computing software, such as through utilize \texttt{torch.autograd()} in Pytorch, when the penalty parameters $\{\lambda_{(\ell)}\}^{L-1}_{\ell = 2}$ are given. Then, the gradient descent from the back-propagation method \citep{rumelhart1986learning} is applied to update the coefficients across all layers. The DPS fitting procedure is detailed in Algorithm 1 of Appendix E of Supplemental Material.

The implementation of fitting DPS (Algorithm 1 of Appendix E) requires adequate network structure selection. An efficient methods are provided in the next Section with theoretical justification.

\section{Fast Tuning and Network Structure Selection}

\noindent Instead of relying on traditional methods such as cross-validation and exhaustive grid search for network structure selection, we propose a novel framework that can optimize penalty parameters while simultaneously offering Generalized Cross Validation Scores (GCVs) \citep{golub1979generalized} for selecting the structure of the deep neural network. This approach treats the weight parameters of each hidden layer as random effects, modeled with appropriate prior distributions.  This enables an Expectation-Maximization (EM)-based algorithm \citep{dempster1977maximum, wu1983convergence} can be developed for tuning the penalty parameters and the structure. Furthermore, several of the theoretical properties of the estimators from the proposed algorithm can be established through their connections with theorems from Expected Conditional Maximization (ECM) method \citep{meng1993maximum}.

The method starts from viewing the outputs of the $\ell$-th layer for $\ell = 1, \cdots, L-1$ in DPS model (\ref{eq: nonlin_DL}) is the conditional expectation of the response conditioned on inputs from the following model
\begin{eqnarray}\label{eq: Ind_coef_models}
    \mathbf{y}_{j, (\ell)} & = & h_{j,(\ell)}\left(\mathbf{h}_{(\ell-1)}(\cdots\left(\mathbf{h}_{(1)}(\{\mathbf{x}_i\}_{i=1}^n)\cdots\right)\right) + \boldsymbol{\epsilon}_{i, (\ell)}, \nonumber\\
   \mathbf{y}_{(L)} & = & \mathbf{h}_{(L)}\left(\mathbf{h}_{(L-1)}(\cdots\left(\mathbf{h}_{(1)}(\{\mathbf{x}_i\}_{i=1}^n)\cdots\right)\right) + \boldsymbol{\epsilon}_{(L)}, 
\end{eqnarray}
where $\mathbf{y}_{j,(\ell)}$ is the response of $j$-th neuron in $\ell$-th layer, $\{\boldsymbol{\epsilon}_{j, (\ell)}\}^{p_{(\ell)}}_{j=1}$ and $\boldsymbol{\epsilon}_{(L)}$ are independent noise vectors with mean $\mathbf{0}$ and variance $\sigma^{2}_{(\ell)} \mathbf{I}_n$, where $ \mathbf{I}_n$ is an $n \times n$ identity matrix. These noises accounts for approximation errors in each layer. Recall that from (\ref{eq: nonlin_DL}), each function $\mathbf{h}_{(\ell)}$ depends on the weights $\mathbf{W}_{(\ell)}$ for $\ell = 1, \cdots, L-1$ and $\mathbf{w}_{(L)}$ for layer $L$. If $\{\mathbf{W}_{(\ell)}\}^{L-1}_{\ell = 1}$ and $\mathbf{w}_L$ are treated as independent random effects across layers with output of each layer denoted by $\mathbf{y}_{(\ell)} = (\mathbf{y}_{1, (\ell)}, \cdots, \mathbf{y}_{p_{(\ell)}, (\ell)})$ for layer $\ell = 1, ... L$, the likelihood function of (\ref{eq: Ind_coef_models})  given input data can be expressed as:
\begin{eqnarray} \label{eq: Like_Decom}
 & & \prod^{p_{(L)}}_{j=1} f(\mathbf{y}_{j,(L)}, \{\mathbf{y}_{(\ell)}\}^{L-1}_{\ell = 1},  \{\mathbf{W}_{(\ell)}\}^{L-1}_{\ell = 1},\mathbf{w}_L  \mid  \{\mathbf{x}_{i}\}^{n}_{i = 1}) f(\{\mathbf{W}_{(\ell)}\}_{\ell=1}^{L-1},\mathbf{w}_L)  \nonumber\\
 & = & \prod^{p_{(L)}}_{j=1}\ f\left(\mathbf{y}_{j,(L)} \mid  \{\mathbf{x}_{i}\}^{n}_{i = 1}, \{\mathbf{y}_{(\ell)}\}^{L-1}_{\ell = 1}, \{\mathbf{w}_{(L)}\} \right)  f\left(\{\mathbf{y}_{(\ell)}\}^{L-1}_{\ell = 1}, \{\mathbf{W}_{(\ell)}\}^{L-1}_{\ell = 1} \mid  \{\mathbf{x}_{i}\}^{n}_{i = 1}\right ) f(\{\mathbf{W}_{(\ell)}\}_{\ell=1}^{L-1},\mathbf{w}_L)\nonumber \\
    & = & \underbrace{\prod^{p_{(L)}}_{j=1} f\left(\mathbf{y}_{j,(L)}\mid  \{\mathbf{x}_{i}\}^{n}_{i = 1}, \{\mathbf{y}_{(\ell)}\}^{L-1}_{\ell = 1}, \{\mathbf{w}_{(L)}\} \right)f(\mathbf{w}_{(L)})}_{\text{Last layer likelihood}}
     f\left (\{\mathbf{y}_{(\ell)}\}^{L-1}_{\ell = 1}, \{\mathbf{W}_{(\ell)}\}^{L-1}_{\ell = 1} \mid  \{\mathbf{x}_{i}\}^{n}_{i = 1}\right )  f(\{\mathbf{W}_{(\ell)}\}_{\ell=1}^{L-1})\nonumber\\
        & = & \underbrace{\prod^{p_{(L)}}_{j=1}\left\{f\left(\mathbf{y}_{j,(L)}\mid  \{\mathbf{x}_{i}\}^{n}_{i = 1}, \{\mathbf{y}_{(\ell)}\}^{L-1}_{\ell = 1}, \{\mathbf{w}_{(L)}\} \right)f(\mathbf{w}_{(L)})\right\}}_{\text{Last layer likelihood}} \nonumber \\
    &\times & \underbrace{\prod^{p_{(L-1)}}_{k=1}\left\{f(\mathbf{y}_{k,(L-1)} \mid  \{\mathbf{x}_{i}\}^{n}_{i = 1}, \{\mathbf{y}_{(\ell)}\}^{L-2}_{\ell = 1}, \{\mathbf{W}_{(L-1)}\})f(\mathbf{W}_{(L-1)})\right\}}_{\text{Second Last layer likelihood}} \nonumber \\
    & \times &f(\{\mathbf{y}_{(\ell)}\}^{L-2}_{\ell = 1}, \{\mathbf{W}_{(\ell)}\}^{L-2}_{\ell = 1} \mid  \{\mathbf{x}_{i}\}^{n}_{i = 1}) \times f(\{\mathbf{W}_{(\ell)}\}_{\ell=1}^{L-2})\nonumber\\
    & = & \cdots \nonumber\\
    & = & \prod_{\ell=1}^L\ \ell\text{-th layer likelihood},
\end{eqnarray}
where $f(\cdot)$ is the density function of its associated parameter. For instance, $f\left(\{\mathbf{W}_{(\ell)}\}^{L-1}_{\ell = 1},\mathbf{w}_{(L)}\right)$ represents the joint prior density function of the random effects $\{\mathbf{W}_{(\ell)}\}^{L-1}_{\ell = 1}$ and $\mathbf{w}_{(L)}$ of the network. Selecting appropriate prior distributions for $f(\mathbf{W}_{(\ell)})$ and $f(\mathbf{w}_L)$ enables closed-form update for hyperparameters, as detailed in the following theorem. 



\begin{theorem}\label{thm: ECM}
Based on the model described by (\ref{eq: Ind_coef_models}) with the layer-wise independence assumption for the likelihood function (\ref{eq: Like_Decom}), and the regression effects from each neuron in $\ell$-th layer are assumed to follow normal distribution as: 
$$(\boldsymbol{\omega}_{1,(\ell)}, \cdots, \boldsymbol{\omega}_{p_{(\ell)},(\ell)}) \overset{iid}{\sim} \mathcal{N}(0, \xi^2_{(\ell)} \mathbf{S}^{-1}_{(\ell)}).$$ 
Here, \(\mathbf{S}_{(\ell)}\) is defined as \(\mathbf{D}_{(\ell), r} \mathbf{D}^{T}_{(\ell),r}\), where $\mathbf{D}_{(\ell),r}$ represents the $r$-th order difference matrix. The parameter \(\xi^2_{(\ell)}\) denotes the variance of the random regression effects in $\ell$-th layer. Additionally, $\epsilon_{(\ell)} \overset{iid}{\sim} \mathcal{N}(0, \sigma_{(\ell)}^2 \mathbf{I}_n)$ from (\ref{eq: Ind_coef_models}). Under these conditions, we can derive the following estimators that maximize the likelihood function \((\ref{eq: Like_Decom})\).


\begin{itemize}
    \item[(a)] Based on the observed data, each neuron in first layer, $\ell=1$, can be interpreted as a multi-linear model. Consequently, the weights for the $j$-th neuron ($\boldsymbol{\omega}_{i,(1)}$) can be estimated using ordinary least squares:
    $$
    \widehat{\boldsymbol{\omega}}_{j,(1)}=(\mathbf{X}^T\mathbf{X})^{-1}\mathbf{X}^T\mathbf{y}_{j,(1)},\mbox{ for }1\leq j\leq p_{(\ell)},
    $$
    where $\widehat{\boldsymbol{\omega}}_{j,(1)}$ is the updated weights for $j$-th row of weight matrix $\mathbf{W}_{(1)}$.
    
    \item[(b)] Given the observed data, assuming known value for $\sigma_{(\ell)}^2$ and $\xi^2_{(\ell)}$, the estimator for the variance of the random effects, \(\hat{\xi}_{(\ell)}^2\), can be computed as:
    \[
    \hat{\xi}_{(\ell)}^2 = \frac{1}{p_{(\ell)}}  \sum_{j=1}^{p_{(\ell)}} 
    \left( \operatorname{Tr}(\mathbf{S}_{(\ell)} \Gamma_{j}^{(\ell)}) 
    + \left( \mu_{j}^{(\ell)} \right)^T \mathbf{S}_{(\ell)} \mu_{j}^{(\ell)} \right),
    \]
    where \(\mu_{j}^{(\ell)}\) and \(\Gamma_{j}^{(\ell)}\) are the conditional mean and covariance of the random effects in layer \(\ell\) given the observed data.

    \item[(c)] Given the observed data, assuming known value for $\sigma_{(\ell)}^2$ and $\xi^2_{(\ell)}$, the estimator for the variance of the residuals, \(\hat{\sigma}_{(\ell)}^2\), can be constructed as:
    \[
\hat{\sigma}_{(\ell)}^2 = \frac{1}{n p_{(\ell)}} \sum_{j=1}^{p_{(\ell)}} \left( \| \mathbf{y}_{j,(\ell)} \|^2 - 2 (\mathbf{y}_{j,(\ell)})^T \boldsymbol{\omega}_{j,(\ell)} \mu_{j}^{(\ell)} + \operatorname{Tr}\left((\boldsymbol{\omega}_{j,(\ell)})^T \boldsymbol{\omega}_{j,(\ell)} \Gamma_{j}^{(\ell)}\right) + \left( \mu_{j}^{(\ell)} \right)^T (\boldsymbol{\omega}_{j,(\ell)})^T \boldsymbol{\omega}_{j,(\ell)} \mu_{j}^{(\ell)} \right).
\]

    \item[(d)] Given the observed data, the estimator for the regularization parameter \(\hat{\lambda}_{(\ell)}\), assuming known values of $\hat{\xi}_{(\ell)}^2$ and $\hat{\sigma}_{(\ell)}^2$, is estimated as:
    \[
    \hat{\lambda}_{(\ell)} = \frac{\hat{\sigma}_{(\ell)}^2}{\hat{\xi}_{(\ell)}^2}.
    \]

    \item[(e)] For the last layer $L$, the estimates of $\mathbf{w}_{(L)}$ can be calculated through multivariate response regression:
    $$
    \widehat{\mathbf{w}}_{(L)}=\left(\mathbf{H}_{(L-1)}^T\mathbf{H}_{(L-1)}\right)^{-1}\mathbf{H}_{(L-1)}^T\mathbf{y}_{(L)},
    $$
    where $\mathbf{H}_{(L-1)}$ is the model matrix from the second last layer, whose $j$-th row is $\mathbf{y}_{j, (L-1)}$ for $j = 1, \cdots, p_{(L - 1)}.$   
\end{itemize}
\end{theorem}
The proof of this theorem is provided in Appendix B of Supplemental Material.  The theorem provides closed-form solutions of the expected conditional steps of an ECM algorithm for iteratively updating the estimators of parameters in DPS, with $\hat{\lambda}_{(\ell)}$ in each layer being adjusted per iteration. Algorithm 2 in Appendix E outlines the complete tuning procedure.




Algorithm 2 in Appendix E benefits from the robust convergence properties of ECM algorithm. The assumption of layer-wise independence aligns with the convergence criteria established proposed by \citep{meng1993maximum}, ensuring stable and reliable parameter estimates. While this framework employs specific priors, it is extendable to accommodate alternative prior choices, enabling broader Bayesian interpretations. For our selected prior, the proposed methodology provides closed-form updates for parameter estimation during each iteration; Thus, the proposed method enjoys both computational efficiency benefits from the closed form solution and rapid convergence from the theories of the ECM algorithm.

Additionally, incorporating the chosen prior distribution leads to additional advantages of the network structure selection. The key relies on the closed form solution and its connection with smoothing matrix in nonparametric regression literature \citep{hardle1990applied}. Specifically, the coefficient estimator of all coefficients of the last layer in Results (5) of Theorem 3.1 motivates us considering the projection matrix $\mathbf{P}_{(L)} = \mathbf{H}_{(L-1)} (\mathbf{H}^{T}_{(L-1)}\mathbf{H}_{(L-1)})\mathbf{H}^{T}_{(L-1)}$. The summation of the diagonal elements in $\mathbf{P}_{(L-1)}$ denoted $trace(\mathbf{H}_{(L-1)})$ represents the effective number of parameters in the model as its dependency of all previous layers.  By calculating the trace of the projection matrix, analogous to computing degrees of freedom in smoothing matrices, and hence a GCV criterion under current model $\mathcal{M}$ can be obtained by (\ref{eq: GCV}).
\begin{equation}\label{eq: GCV}
    \text{GCV}(\{\mathbf{x}_i\}_{i=1}^n;\mathcal{M}) = \frac{\sum_{i=1}^n \left(\mathbf{y}_i - \hat{f}(\mathbf{x}_i)\right)^2}{\left(n - \text{trace}(\mathbf{P}_{(L-1)})\right)^2},
\end{equation}
This formulation enables meaningful comparisons across models with varying structures, guiding the selection of the model that minimizes the GCV score. The chosen model achieves an optimal balance between predictive accuracy and model complexity. The complete tuning procedure is outlined in Algorithm 3 in Appendix E.

We rigorously compare the proposed algorithm to conventional backpropagation with gradient descent for DNNs. Suppose there are \( L \) layers, each with \( M \) linear neurons. The computational cost of updating coefficients in a layer using gradient descent is \( O(M^3) \), making the total complexity for \( L \) layers with \( T \) iterations \( O(LTM^3) \). With \( n \)-fold cross-validation and \( M \), \( L \), and \( T \) each scaling with \( n \), the complexity is \( O(n^6) \). On the other hand, our fast-tuning method, by Theorem 1, shows that \( n \) linear neurons are equivalent to \( M/N \) spline neurons with \( N \) knots. By Theorem 3, the complexity per layer is \( O((M/N \cdot N)^3) \), giving an overall complexity of \( O(EL(M/N \cdot N)^3) \), where \( E \) is the number of ECM algorithm iterations. With \( L \) and \( M \) scaling with \( n \), the fast-tuning complexity is \( O(En^4) \), significantly faster than conventional methods as \( E \) is typically small. To validate, Algorithm 3 in Appendix E is tested with a training dataset with sample size 800 and extra 200 testing data points from the model
\[
y = \sin(x_{1}) + \cos(x_{2}) + \epsilon,
\]
where $\epsilon$ is noise whose mean is 0 and variance is chosen for making its value to be  $5\%$ of the sample variances of values  $\sin(x_{1}) + \cos(x_{2})$ evaluated at training dataset. We set the number of parameters from the network structure candidates of DPS to be similar to that from DNN for creating a fair comparison environments: for DNN, 8 candidates of neuron number are considered $M \in\{240,260,\cdots,360\}$ and for DPS 8 candidates of neuron number and knot numbers are considered $p_{(\ell)} \in \{10, 20, 30, 40\}$ with $N \in \{10, 15\}$, so the total number of unknown parameters are around 720 to 1500 for one layer.  The resulting running time of DPS is 43.112 seconds with MSPE 0.0122, while the running time of DNN is 125.183 with MSPE 0.0268. The results support our method enjoys faster network structure selection and improves predictive performance of the proposed method over standard DNNs.

\section{Numerical Studies}\label{sec:sim}
In this section,  more simulation studies conducted for observing the performance of DPS. In section 4.1, we compare DPS with other network method including conventional deep (linear) neural network denoted by DNN  \citep{schmidhuber2015deep, lecun2015deep}. In section 4.2, we further compare with existing nonparametric regression methods including   multivariate adaptive regression splines denoted by MARS \citep{friedman1991multivariate}) and P-Spline
\citep{eilers2021practical}. For all the methods we compared,  5-fold cross validation is implemented for selecting their optimal structure; for DPS, we implement Algorithm 3 in Appendix E for its structure selection. Additionally, the fact that DPS includes the difference penalty to automatically determine model complexity motivates us to explore if the proposed method can overcome model selection problems from double descent phenomenon observed in many machine learning models. More details on this direction are given in Section 4.3.




\subsection{Numerical Comparisons with Other Network Methods}\label{subsec:ex1}
The models are trained on data generated from the following true function:
\begin{equation}\label{eq:s1}
    y = \exp\{2\sin(0.5\pi x_1) + 0.5\cos(2.5\pi x_2)\} +\epsilon 
\end{equation}
where $\mathbf{x} = (x_1, x_2) \in [0, 1]^2$ and $\epsilon \sim\mathcal{N}(0,\sigma^2)$ where $\sigma^2$ is chosen so that the error variance is $5\%$ of the sample variance of function values $\exp\{2\sin(0.5\pi x_1) + 0.5\cos(2.5\pi x_2)\}$ evaluated at the input data. The model \ref{eq:s1} under consideration involves a two dimensional input vector where $x_1 \sim U(0, 1)$ and $x_2 \sim U(0, 1)$ independently. The study examines three sample sizes: $n \in \{200, 400, 800\}$, with additional independent testing set of 500 samples generated from the same input distributions. To ensure robustness, the simulation setting is repeated 100 times for testing the robustness of the results, and the averages (and standard deviations) of the mean squared prediction error (MSPE) of the 100 simulations  are recorded and summarized in Table \ref{tab:RMSE1}. The considered network structure includes the number of layers $L$ ranging from 1 to 4, and the number of neurons per layer $n \in \{50, 60, 70, 80 ,90 , 100\}$ and the two-layer B-Spline network (2DS) with 50 neurons, 15 knots, and the third-order basis functions. The optimal network structure is determined using 5-fold cross-validation. The results indicate that the two-layer P-Spline network (2DPS) with 50 neurons, 15 knots, and the third-order basis functions outperforms other methods across all scenarios considered.



\begin{table}[]
\centering
\begin{tabular}{c|ccc}
\hline
\multicolumn{1}{c|}{Method} & \multicolumn{1}{c|}{2DS} & \multicolumn{1}{c|}{2DPS}  & \multicolumn{1}{c}{DNN}\\ \hline
\multicolumn{1}{c|}{Training Size}   & \multicolumn{3}{c}{MSPE} \\ \hline
\multicolumn{1}{c|}{N=200}  & \multicolumn{1}{c|}{0.088 (0.081)} & \multicolumn{1}{c|}{\textbf{0.051} (0.031)}   & 0.585 (0.484) \\ \hline
\multicolumn{1}{c|}{N=400}  & \multicolumn{1}{c|}{0.078 (0.217)} & \multicolumn{1}{c|}{\textbf{0.028} (0.008) }    &  0.078 (0.155)  \\ \hline 
\multicolumn{1}{c|}{N=800}  & \multicolumn{1}{c|}{0.044 (0.076)} & \multicolumn{1}{c|}{0.024 (0.008)}   &  \textbf{0.014} (0.012) \\ \hline
\end{tabular}
\caption{Average (sd) of MSPEs for Example 1}
\label{tab:RMSE1}
\end{table}

\subsection{Comparisions with Other Nonparametric Regression Methods}
In this subsection, three more simulation functions
\begin{align*}\label{eq:s2}
    g^*_1(\textbf{x})&= \left[\prod^p_{i=1}\frac{|4x_i-2|+a_i}{1+a_i}\right],\text{ where }a_i=i/2,i=1,\cdots,p\\
    g^*_2(\textbf{x})&= \sum_{i=1}^p(-1)^{i}i^{-1}x_i\\
    g^*_3(\textbf{x})&= \left[\exp\left(x_1^2-\sqrt{x_2+5}\right)+0.01\cot\left(\frac{1}{0.01+|x_3+x_4|}\right)\right]
\end{align*}
and two more nonparametric regression methods, MARS and P-Spline,  are considered.  We first independently generate $p$-dimensional input $\textbf{x}$ from uniform distributions on the interval $[0,1]^p$ and generate the corresponding the results $\tilde{g}^*_{\gamma}(\textbf{x})$ from noise models: $\tilde{g}^*_{\gamma}(\textbf{x})= g^*_{\gamma}(\textbf{x})+\epsilon,$ where $\gamma = 1, 2, 3$ and $\epsilon$ is a Gaussian noise with mean 0 and the variance $\sigma^2$ are chosen so that the error variance becomes $5\%$ of the sample variance of the response from the true function (${g}^*_{\gamma}(\textbf{x})$). Subsequently, we compare the performance between proposed model DPS and others with respect to training size $N=800$ and testing size $n=400$. The mean squared prediction error (MSPE) from the fitted models are given in Table \ref{tab:RMSE3}, demonstrating the great performance of the proposed model compared with other methods. For comparative analysis, we also applied the MARS method to these examples and the results are summarized in Table \ref{tab:RMSE3}. Specifically, our proposed penalized model achieved comparable or superior results to MARS for example $\tilde{g}^*_{2}$ and $\tilde{g}^*_{3}$. Besides, compared to the P-Spline method, we can observe that our proposed method have better performance for almost all examples. This outcome underscores the efficiency and effectiveness of our approach in optimizing neural network performance.

The results presented in Table \ref{tab:RMSE3} demonstrate the efficacy of GCV in model selection. Following the selection process, the chosen models underwent with a fast tuning procedure involving layer-wise weight updates with a difference penalty. Notably, for the second and third examples, our proposed penalized model achieved comparable or superior results to MARS, without the need for cross validation in hyperparameter selection. Besides, compared to the traditional P-Spline, we can observe that our proposed method have better performance by adding constraints to the basis expansion on neurons in the network. This outcome underscores the efficiency and effectiveness of our approach in optimizing neural network performance.

\begin{table}[]
\centering
\resizebox{16cm}{!}{
\begin{tabular}{ccccccccccc}
\hline
\multirow{3}{*}{True function} & \multicolumn{10}{c}{Sample size}   \\ \cline{2-11} 
    & \multicolumn{5}{c}{N = 400}                         & \multicolumn{5}{c}{N = 800} \\ \cline{2-11} 
    & \multicolumn{1}{c}{DPS} &  \multicolumn{1}{c}{DS} & \multicolumn{1}{c}{P-Spline} & \multicolumn{1}{c}{DNN} & MARS & \multicolumn{1}{c}{DPS} &  \multicolumn{1}{c}{DS} & \multicolumn{1}{c}{P-Spline} & \multicolumn{1}{c}{DNN} & MARS \\ \hline
$g^*_1\ (p=2)$ & \multicolumn{1}{c}{0.0016}    & \multicolumn{1}{c}{0.0058} & \multicolumn{1}{c}{0.0137}   & \multicolumn{1}{c}{0.0153}   &   {\bf 0.0003}   &  \multicolumn{1}{|c}{0.0013}  & \multicolumn{1}{c}{0.0037} & \multicolumn{1}{c}{0.0135}     & \multicolumn{1}{c}{0.0117}   &  {\bf 0.0004}  \\

$g^*_2\ (p=2)$  &   \multicolumn{1}{c}{\textbf{<0.0001}}    & \multicolumn{1}{c}{0.0008} &  \multicolumn{1}{c}{0.0001}  & \multicolumn{1}{c}{0.0028}   &  0.0001   & \multicolumn{1}{|c}{\textbf{<0.0001}}    & \multicolumn{1}{c}{0.0009} &  \multicolumn{1}{c}{0.0001}    & \multicolumn{1}{c}{0.0019}   &  0.0001    \\  

$g^*_3\ (p=4)$  & \multicolumn{1}{c}{\textbf{<0.0001}}   & \multicolumn{1}{c}{0.0014} &  \multicolumn{1}{c}{0.0042}    & \multicolumn{1}{c}{0.0013}   &  0.0001  & \multicolumn{1}{|c}{\textbf{<0.0001}}    & \multicolumn{1}{c}{0.0009} & \multicolumn{1}{c}{0.0001}    & \multicolumn{1}{c}{0.0014}   &   0.0001   \\ \hline
\end{tabular}}
\caption{MSPE of Example 2}
\label{tab:RMSE3}
\end{table}

\subsection{Overcome Model Selection Problems of Double Descent}
Double descent is a surprising phenomenon observed in many machine learning models: it is well known that the test error of these models initially decreases, then increases as the number of model parameters grows; however, as the number of parameters continues to increase into the highly overparameterized regime, the test error decreases again \citep{nakkiran2020optimal,nakkiran2021deep}. Although the test error eventually drops, previous studies do not provide a clear answer as to which model one should select: the model from the second descent (overparameterized), one from an underparameterized model, or others. An advantage of the proposed DPS is that its difference penalty can smooth the effect of closely spaced knots, making knot selection less critical and thus addressing the double descent dilemma in deep regression models.

\begin{figure}
\includegraphics[width=1.0\linewidth]{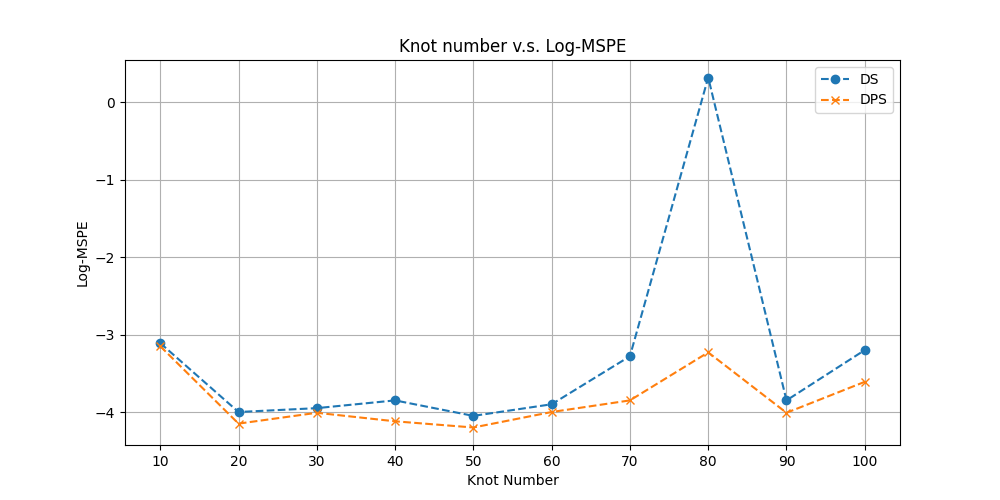}
\caption{Testing MSPE over DPS and DS with respect to various knots}
\label{fig:doubleDPS}%
\end{figure}

To demonstrate this advantage, we conduct simulation experiments using DS and DPS under the same conditions outlined in Section \ref{subsec:ex1}, with a training set of 800 samples and a testing set of 400 samples. The resulting MSPE, in logarithmic scale, for both methods is summarized in Figure \ref{fig:doubleDPS}. For DS, the test error initially decreases, then rises as the model becomes overparameterized. However, this phenomenon is attenuated in DPS, indicating that its performance is less sensitive to the overspecification of model parameters. Specifically, DPS exhibits consistent performance across a range of knot numbers from 20 to 100. Additionally, under the same number of knots, the result demonstrates that DPS consistently outperforms DS, aligning with the conclusions from the previous section.


\section{A Convergence Theorem of DPS}
This section explores a general convergence theorem of DPS for model (\ref{eq: nonlin_DL})  with the independent layer coefficient assumption as a special. 

\subsection{Convergence Rates of Mean Square Prediction Errors}
The model class of (\ref{eq: nonlin_DL})  can be described as (p, C)-smooth hierarchical composition models, defined as below:

\begin{definition}\label{def: PCsmooth}
 A function \(m: \mathbb{R}^{d} \rightarrow \mathbb{R}\) is called \((p, C)\)-smooth if \(p = q + s\) for some \(q \in \mathbb{N}_{0}\) and \(0 < s \leq 1\) and for every $\boldsymbol{\alpha}=\left(\alpha_{1}, \ldots, \alpha_{d}\right) \in \mathbb{N}_{0}^{d}$ with $\sum_{j=1}^{d} \alpha_{j}=q$ the partial derivative $\partial^{q} m /\left(\partial x_{1}^{\alpha_{1}} \ldots \partial x_{d}^{\alpha_{d}}\right)$ exists and satisfies

$$
\left|\frac{\partial^{q} m}{\partial x_{1}^{\alpha_{1}} \ldots \partial x_{d}^{\alpha_{d}}}(\mathbf{x})-\frac{\partial^{q} m}{\partial x_{1}^{\alpha_{1}} \ldots \partial x_{d}^{\alpha_{d}}}(\mathbf{z})\right| \leq C\|\mathbf{x}-\mathbf{z}\|^{s}
$$

for all $\mathbf{x}, \mathbf{z} \in \mathbb{R}^{d}$, where $\|\cdot\|$ denotes the Euclidean norm.\\
\end{definition}

\begin{definition}\label{def: HierCompM}
A DNN model with network structure $\mathcal{N} = (L, \bm{p})$, where $\bm{p} = \{\bm{p}_{(1)}, \ldots, \bm{p}_{(L)}\} \in \mathbb{N}^{L}$ specifies the number of neurons in each layer and $L$ represents the number of layers is called a (p, C)-smooth hierarchical composition models if the following conditions are satisfied

\begin{itemize}
    \item[(a).] For the input layer, if there exists $\mathbf{a}_{1}, \ldots, \mathbf{a}_{p_{(1)}} \in \mathbb{R}^{d}$ and the function $f: \mathbb{R}^{p_{(1)}} \rightarrow \mathbb{R}$ combines all neurons in this layer 
$$
h(\mathbf{x})=f\left(\mathbf{a}_{1}^{\top} \mathbf{x}, \ldots, \mathbf{a}_{d^{*}}^{\top} \mathbf{x}\right) \quad \text { for all } \mathbf{x} \in \mathbb{R}^{d}
$$
is (p, C)-smoothness. 
\item[(b).] For $\ell$-th layer from $\ell = 2, \cdots, L$, if there exists $ g_{k}: \mathbb{R}^{p_{(\ell - 1)}} \rightarrow \mathbb{R}$ for $k \in \{1, \ldots, p_{(l-1)}\}$ and $f_{1, k}, \ldots, f_{p_{(\ell-1)}, k}: \mathbb{R}^{d} \rightarrow \mathbb{R}$ for $k \in\{1, \ldots, p_{(l-1)}\}$ such that all functions $f$ and $g_{k}$ and
$$
h(\mathbf{x})=\sum_{k=1}^{p_{(\ell)}} g_{k}\left(f_{1, k}(\mathbf{x}), \ldots, f_{p_{(\ell-1)}, k}(\mathbf{x})\right) \quad \text { for all } \mathbf{x} \in \mathbb{R}^{d}
$$
are $(p, C)$-smooth.
\end{itemize}
    
\end{definition}


Under the model class and mild conditions listed in Appendix C,  the expected mean square errors from Algorithm 1 can be shown to have the following convergence rates:

\begin{theorem}[Convergence Rate for DPS]
\label{thm:convergence_rate}
Let \( (\mathbf{X}, Y), (\mathbf{X}_1, Y_1), \ldots, (\mathbf{X}_n, Y_n) \) be independent and identically distributed (i.i.d.) random regression paired variables. Then under conditions (i)-(iv) listed Appendix C, the expected \(L_2\) error of the estimator $\hat{\boldsymbol{h}}_{(L)}$ obtained from Algorithm 1 for model (\ref{eq: nonlin_DL}) satisfies:

\[
\mathbf{E} \int \left|\hat{\boldsymbol{h}}_{(L)}(\mathbf{x}) - \boldsymbol{h}(\mathbf{x})\right|^2 \mathbf{P}_\mathbf{X}(d\mathbf{x}) =  O_{p}\left(N (\log n)^{5} n^{-\frac{2p}{2p+K_{eff}}}\right),
\]

for sufficiently large \(n\), where $\boldsymbol{h}$ is the true function belonging to (p, C)-smooth hierarchical composition model space.
\end{theorem}
The proof extensively use the approximation theory of empirical process, for example, \cite{gyorfi2006distribution} and \cite{geer2000empirical}, and a sharp upper bound of VC dimension applied to model (\ref{eq: nonlin_DL}) from \cite{bartlett2019nearly}. The details are summarized in Appendix D in the Supplemental Material.

\subsection{Overcome Curse of Dimensionality of Non-Parametric Regression Methods}
The convergence result established in Theorem \ref{thm:convergence_rate} demonstrates an essential advantage of the proposed framework by mitigating the curse of dimensionality, a common challenge in non-parametric modeling. The curse of dimensionality refers to the exponential growth in the amount of data required to achieve a given level of accuracy as the dimensionality of the input space increases \citep{donoho2000high}. In traditional non-parametric models, such as kernel-based methods, the convergence rate deteriorates significantly as the input dimension grows, making these methods impractical for high-dimensional applications \citep{Tsybakov2009nonparametric, gyorfi2006distribution}.

Unlike conventional nonparametric regression, the convergence rate derived in Theorem \ref{thm:convergence_rate} depends primarily on the effective smoothness parameter $p$ and the penalized effective dimension $K_{eff}$, rather than the raw dimensionality of the input space. This independence from the number of input variables suggests that the deep penalty spline (DPS) framework can adapt to high-dimensional data without suffering from the typical rate degradation observed in non-parametric methods, which depends on input dimension \citep{stone1980optimal}.

This result aligns with the recent literature on neural networks, which highlights their ability to bypass the curse of dimensionality in certain settings \citep{barron1993universal, bach2017equivalence}. The DPS framework, by leveraging hierarchical neural architectures and appropriate penalization on the basis coefficients, achieves similar benefits. The use of penalized B-splines introduces additional flexibility, allowing the model to concentrate complexity where it is most needed, rather than uniformly across all input dimensions.


Except for theoretical justification, we further support the results by numerical studies.  Let us still use $g^*_1$ to generate simulated data with respect to different training size and input dimension that $N=\{200, 400, 800, 1600\}$ and $p={2, 6, 10}$, and summarize the result of the experiment in the Table \ref{tab:RMSE4}. Both DS and DPS use 50 neurons and 15 knots. Table \ref{tab:RMSE4} demonstrates the MSPE results of examples from  $g^*_1$ function with input dimension $d = 2, 6, and 10$ and training dataset size $n = 200, 400, 800, 1600$. From the table, we observe that as sample size $n$ becomes larger, the MSPE values are decreasing. Also, if we count the ratios of MSPE from a larger sample size to a smaller sample size, we see the ratios are closed $80\%$ for d = 2 case, $65\%$ for d = 6 case, $70\%$ for d = 10 case, which support the finding of Theorem 5.1 to have a fixed convergent rate not dependent on the input dimensions.  Such phenomenon is not observed from the results of the DS method in Table  \ref{tab:RMSE4}.

\begin{table}[]
\centering
\begin{tabular}{|cc|c|c|c|}
\hline
\multicolumn{2}{|c|}{MSPE}                        &  n=400 & n=800 & n=1600 \\ \hline
\multicolumn{1}{|c|}{\multirow{2}{*}{d=2}}  & DS    &  0.0042 (0.0007)    &  0.0026 (0.0005)     &    0.0027 (0.0009)     \\ \cline{2-5} 
\multicolumn{1}{|c|}{}                      & DPS &   \textbf{0.0018} (0.0004) & \textbf{0.0015} (0.0003)  &   \textbf{0.0011} (0.0004)    \\ \hline
\multicolumn{1}{|c|}{\multirow{2}{*}{d=6}}  & DS & 0.2236 (0.0518)     & 0.1748 (0.0171)  &   0.1231 (0.0110)     \\ \cline{2-5} 
\multicolumn{1}{|c|}{}    & DPS &    \textbf{0.1815} (0.0202)    & \textbf{0.1452} (0.0150)      &  \textbf{0.0864} (0.0089)     \\ \hline
\multicolumn{1}{|c|}{\multirow{2}{*}{d=10}} & DS &  0.6416 (0.1139)   &  0.4390 (0.0722)  &  0.3104 (0.0368)   \\ \cline{2-5} 
\multicolumn{1}{|c|}{}                      & DPS &   \textbf{0.5431} (0.0899)    &  \textbf{0.3451} (0.0604)  &    \textbf{0.2766} (0.0364)    \\ \hline
\end{tabular}

\caption{MSPE of example $g_1^*(\textbf{x})$ with input dimension $d = 2, 6, and 10$ and training dataset size $n = 200, 400, 800, 1600$}
\label{tab:RMSE4}
\end{table}

\section{Real data analysis}
The convergent rate properties pave the way for the DPS method applied to regression problems with many inputs, in which situation the conventional nonparametric regression methods may not perform well. Two examples are illustrated in the following two subsections.

\subsection{Image data analysis: Brain Tumor}\label{data:brain}
The brain tumor is a mass of abnormal cells within the brain. Unlike other tumors, brain tumors are unique challenges for detection and diagnosis due to their location within the skull and the potential for deep growth within brain tissue \citep{amin2022brain,khalighi2024artificial}. Brain tumors can be classified into two types: malignant and benign with significant differences in the level of danger between them. Early detection of brain tumors is crucial, as timely diagnosis can lead to more effective treatments and improved patient outcomes. This makes brain tumor detection an essential medical imaging task. To be more specific, our task is to classify glioma brain images and normal brain images.

Glioma is a type of brain tumor that originates from glial cells, which are supportive cells in the brain and spinal cord. These tumors are the most common primary brain tumors in adults. Some studies indicate that the exact cause pf glioma is not fully understood, but research has identified genetic mutations may contribution to their development. In this application, we applied the proposed DPS method with its fast tuning algorithm to select the best DPS structure for classifying a brain image is glioma or not. We utilize an open source dataset available on Kaggle \citep{msoud_nickparvar_2021} and randomly select 500 images as a training dataset and 200 images as a testing dataset. These images have been resized to $224\times 224$ pixels each, so we can view the each image data point as a regression data point with an input with dimensions equal to $32\times56\times56$ after being processed by the Convolutional Neuron Networks (CNNs) and a binary output. Demonstration of Glioma brain images, normal brain images, and the resulting images from CNN method are summarized in Appendix F.

The proposed DPS model (\ref{eq: nonlin_DL}) can be used for the classification if the activation function $g(\cdot)$ for the output layer of DPS is the Softmax function. For better training the DPS model, CNNs, the networks widely used for their ability to extract meaningful feature maps from images \citep{lecun1989handwritten, krizhevsky2012imagenet}, are applied to the training dataset for extracting features as our new input variables. Incorporating with CNNs allows use to preserve the power of CNNs for feature extraction while leveraging the capacities of DPS model for classification, potentially leading to improved performance in image classification task.

The optimal model structure of DPS is decided by implementing Algorithm 3 in Appendix E to select the penalty parameter $\boldsymbol{\lambda}$, the number of knots, the number of neurons, and the number of layers. To accomplish model selection in classification problem, we utilize the proposed GCV criteria \ref{eq: GCV} which replaces the residual sum of squares with Pearson $\chi^2$, \citep{o1986automatic} for fitting DPS. In our experiment, we consider the number of neuron $n\in\{50, 80, 100, 150\}$, number of knot $\in\{10, 15, 20\}$, and number of spline layer $\in\{1,2\}$. The resulting network structure is two-layer DPS model, configured with 15 knot number and 100 per layer. 
 
The fitted DPS model is tested by the testing dataset and the performance is measured by the zero-one score function, a common criterion used for classification problem. We also compare with DNN in a fair environment by using the same input variables, obtained from CNNs, and by utilizing cross validation to select its optimal network structure. The resulting average zero-one loss values are 0.9600 for the DPS method while 0.9467 for the DNN method. More results are given in Appendix F. Our method achieves a better performance in this classification structure.

\subsection{Surrogate Modeling of Product Lifetime Modeling}

This subsection demonstrates another application of DPS to be able to assessing prediction values and quantifying the prediction uncertainty. This makes the DPS becomes a method of surrogate modeling, a central concept in the analysis of computer experiments \citep{santner2019design, wu2011experiments}. Surrogate modeling techniques use a statistical model to approximate computer models to explore complex system behaviors efficiently. 

As a motivating example, we focus on analyzing the lifetime of electronic devices. These devices may fail due to particularly mechanism, such as {\it front-end gate oxide breakdown (FEOL TDDB)} \citep{yang2017front}. This phenomenon results from the accumulation of traps in the gate oxide region of transistors due to electrical and thermal stresses during device operation. These traps eventually form a conductive path through the gate oxide, causing device failure. The lifetime function of such device component has been described in previous studies \citep{hsu2020extraction} by the following expression:
\begin{equation}\label{eq: lifetime}
    S(t)  = \exp\left(-\left(\frac{t}{\text{A}_{\text{FEOL}} (\text{WL})^{-\frac{1}{\beta}} e^{-\frac{1}{\beta}} V^{a+bT} \exp\left(\frac{cT+d}{T^2}\right) s^{-1}}\right)^\beta\right),
\end{equation}
where input variables in this model include process-dependent constants $\text{A}_{\text{FEOL}}, a, b, c, d$, voltage $V$ and temperature $T$,  width $W$ and length $L$  of the device, the probability of stress $s$, and shape parameter $\beta$ to characterize the failure rate over time.  

While simulating computer experiments is straightforward, assessing prediction uncertainty is challenging and requires advanced regression models, the focus of this study. In this work, we applied {\it MaxPro design} \citep{joseph2015maximum} to generate space-filling designs for all input variables in (\ref{eq: lifetime}). The ranges of each input variable can be found in Appendix G. With the input datasets, the output data can be generated by $\eta=\text{A}_{FEOL}(\text{WL})^{-\frac{1}{\beta}}e^{-\frac{1}{\beta}}V^{a+bT}\exp(\frac{cT+d}{T^2})s^{-1},$ where $\eta$ is defined as $63\%$ of the devices can be expected to have failed because of the generalized Wei-bull model (\ref{eq: lifetime}). With the training dataset, the DPS model (Algorithm 1 with network structure selection Algorithm 3) can be utilized to find the optimal input setting that maximizes the lifetime $\eta$, which is 82.021 for our experiment. Then, following the process of generating the training dataset, we generate a testing dataset independently to evaluate the performance of the proposed method, DPS. The resulting MSPE for the DPS is 0.0807.  We also compare the result with DS, which possess a lightly larger error 0.0809, 

By utilizing the Hierarchical structure defined as in Definition \ref{def: HierCompM}, we can replace the final layer of DPS with a Gaussian process (GP) model and denote it as DPS-G which possesses closed form formula for quantifying prediction mean and variance and served as the fundamental method in many research associated with surrogate modeling \citep{plumlee2018orthogonal, lin2020transformation, joseph2024rational}. This modification enables the model to generate probabilistic predictions and allows for quantifying prediction uncertainty. More mathematical details of the GP modification on DPS is summarized in Appendix G. The modification further tested by the same testing dataset and the performance of prediction accuracy is improved from 0.0807 for the DPS method to 0.0473 for the DPS with GP modification, which is $58\%$ lower of original MSPE from DPS.

\begin{figure}
\ffigbox{\includegraphics[width=0.9\linewidth]{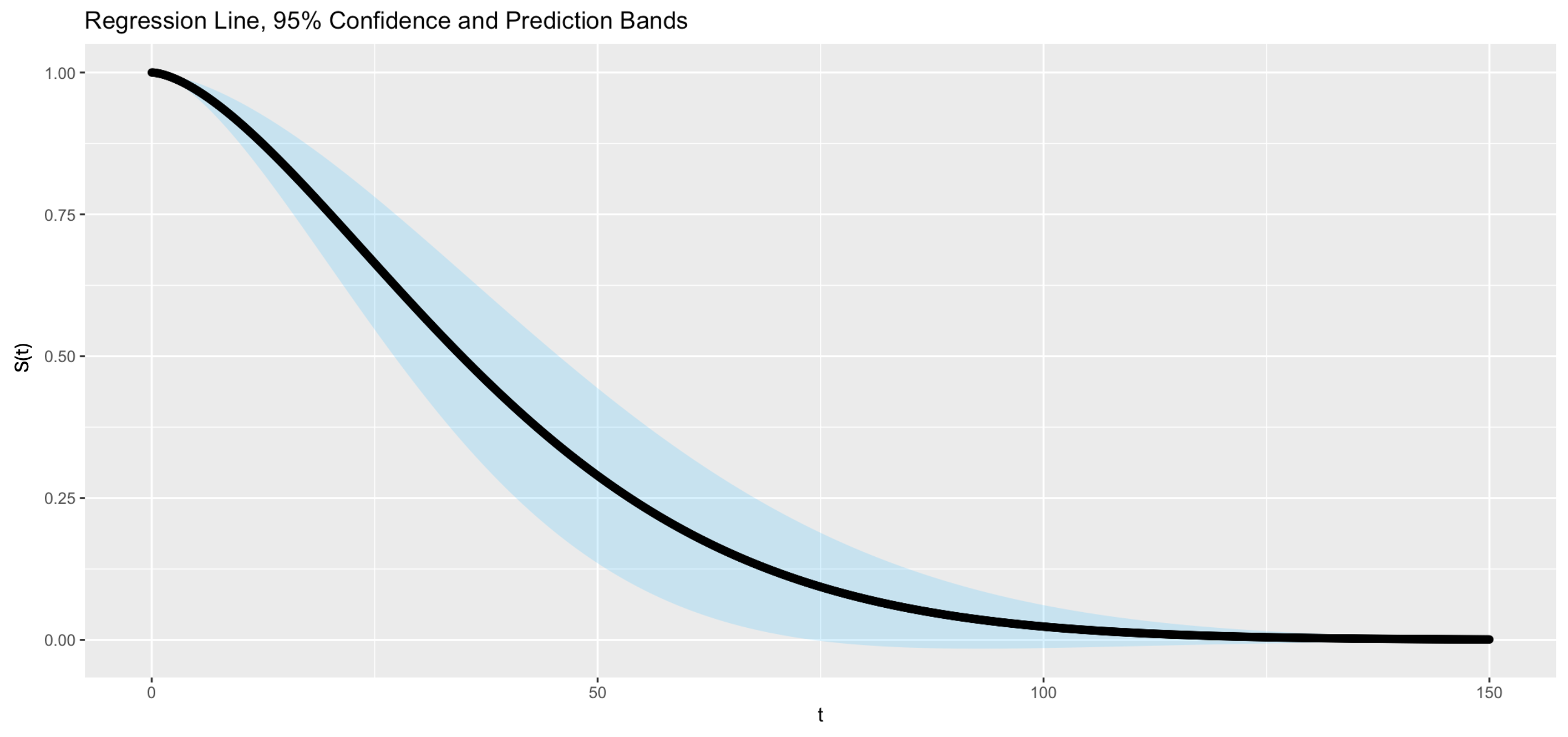}
}{%
  \caption{95 \% confidence interval for the estimated lifetime function (\ref{eq: lifetime})}
  \label{fig:CIPlot2}%
}
\end{figure}

Additionally, together with Delta method \citep{oehlert1992note} and the prediction variance from DPS-G, we are able to quantify prediction uncertainty of $S(t)$. We construct $95\%$ prediction bands over time (Figures \ref{fig:CIPlot2}) based on the optimal setting $\eta=82.021$. From the figure, we observe the interval for the lifetime probability, which is not observed from previous analysis \citep{hsu2020extraction}. Such interval helps us better understand of product quality of the device; for example, we know at $95\%$ confidence more than $75\%$ of devices do not fail after using 25 time units (since the intersection of the under bound curve at time = 25 (the second vertical line from left) and the lifetime probability at 0.75 (second horizontal line from top) is slightly larger than 0.75.  Such information may provide more insights in devise quality and design.

\section{Conclusion and Future Works}\label{sec:dis}
This study introduces a novel regression modeling approach, DPS, which integrates difference penalties with basis expansion techniques for improving deep neural networks (DNNs). The DPS method addresses critical challenges in both DNN modeling and nonparametric regression, delivering significant theoretical and practical benefits. From the DNN modeling perspective, it incorporates an ECM-based algorithm that utilizes the generalized cross-validation (GCV) criterion to efficiently select network structures. This approach streamlines the computationally intensive process of structure selection while maintaining theoretical guarantees. From the perspective of conventional nonparametric regression, the use of difference penalties achieves convergence rates that are independent of input dimensionality. This marks a considerable improvement over conventional nonparametric regression methods, which often struggle with lower convergence rates in high-dimensional regression problems.

Theoretical analyses and comprehensive numerical experiments strongly support the effectiveness of DPS. The versatility of the method is further validated through applications in classification and surrogate modeling, where comparative analyses demonstrate its superior prediction accuracy and interpretability relative to existing methods. These results highlight the potential of DPS as a robust and powerful tool for diverse applications in machine learning and statistical modeling.

Moreover, the DPS framework shows promise for extension to more complex data structures. While current numerical demonstrations focus on scenarios with adequate sample sizes, addressing the challenge of low sample size and high-dimensional (LSS-HD) data emerges as a significant avenue for future research. Developing a robust DPS framework for LSS-HD settings, accompanied by rigorous theoretical properties, represents an exciting and practical direction. Such advancements will further broaden the applicability and enhance the utility of DPS in addressing increasingly complex data analysis scenarios.

\bibliographystyle{agsm}

\bibliography{Ref_all}

\end{document}